\documentclass[a4paper]{mem}
\usepackage{hyperref}
\usepackage{txfonts}
\usepackage{natbib}
\usepackage{graphicx}
\idline{75}{1}
\begin{document}
 \title{WFXT studies of the stellar populations in the Galaxy 
}
 
 \author{S. Sciortino
}
 \offprints{S. Sciortino, sciorti@astropa.inaf.it}

 \institute{INAF-Osservatorio Astronomico di Palermo Giuseppe S. Vaiana,
Piazza del Parlamento 1, 90138 Palermo, Italy,
 \email{sciorti@astropa.inaf.it} \\
 }

\abstract{
I will highlight and discuss some of the studies of the stellar population in 
the Galaxy that will become possible with or will greatly advantage of the
capability of a Wide Field of view X-ray Telescope (WFXT) mission. This mission concept, that was been around for more than 15 years, recently has been re-proposed with renovated interest as part of the US Decadal Astronomy Survey.  

 }
 \authorrunning{S. Sciortino}
 \titlerunning{WFXT studies of the stellar populations in the Galaxy}
 \maketitle
%

\section{Introduction}

The idea of a Wide Field X-ray Telescope mission as one key step for the 
advancement of today astrophysics has been around from nearly 20 years.
Such a mission, conceived by R. Giacconi, has as its main driver the investigation of the large scale structure of the Universe,
by tracing the hot X-ray emitting plasma in the clusters of galaxies
(cf. the Borgani's contribution) and the population of the farthest AGN (cf. the Gilli's contribution). A key element of 
such a mission is a "new-technology" X-ray mirror
in which we trade angular 
resolution near the fov center improving the PSF shape (and resolution) at large off-axis angles over a wide field of view.
(cf. Pareschi's contribution 
for a detailed discussion). The concrete realizability of a WFXT\footnote{In 
1998 it was realized a mirror shell prototype with an almost uniform 10" angular resolution over a $\sim$ 1 sq.deg. fov} has been demonstrated by \citet{1999SPIE.3766..198C}
as part of the study phase of an Italy-USA WFXT mission proposed for the ASI small satellite program \citep{ss_ChiASI98}. 

It was clear, since the beginning, that a WFXT mission 
is a terrific machine for a very ample variety of astrophysical investigations. 
The key characteristics of the mission that has been proposed to the US Decadal Survey are summarized in Table \ref{ss_table1}.
In the following I will try 
to briefly show the WFXT role for studies of stellar populations in the Galaxy. In all the cases 
that I will consider -- a highly selected, personal taste, choice -- a multiwavelength approach is crucial where key WFXT data need to be complemented
with optical, IR, etc. observations. Apart the many possible stellar studies in the Galaxy one can think of, I predict that WFXT observations will results in many new serendipitous exciting discoveries. In this respect let me just mention, as an example, the very recent
XMM-Newton discovery of very intense (100 times higher than expected) X-ray emission from a young brown dwarf \citep{ss_Stelzer2010} that is
urging us to somehow reconsider the X-ray emission mechanism possibly at work in the very low-mass objects.

\begin{table}
\begin{center}
\caption{WFXT Planned Survey Sensitivity}
\begin{tabular} {
|c | c | c | c|
}

\hline
\multicolumn{2}{|c|}{} & \multicolumn{2}{c|}{S$_{min}$ (0.5 - 2 keV)} \\ 
\multicolumn{2}{|c|}{Survey}         & \multicolumn{2}{c|}{point-like at 5$\sigma$} \\
\multicolumn{2}{|c|}{}         & \multicolumn{2}{c|}{erg s$^{-1}$ cm$^{-2}$} \\ 
\hline
\hline
Area & Exposure  & \multicolumn{2}{c|}{Performance} \\
(sq. deg.) & (ksec) & Goal & Baseline\\
\hline
20000 & 2 & 3 x 10$^{-15}$ & 5 x 10$^{-15}$ \\
3000 & 13  & 5 x 10$^{-16}$ & 1 x 10$^{-15}$ \\ 
100 & 400 & 3 x 10$^{-17}$ & 1 x 10$^{-16}$ \\
\hline
\end{tabular}
\end{center}

\small{Values, taken from Rosati's contribution, refer to goal performances (A$_{eff}$~=~1~m$^2$, HEW~=~5"), and to minimal requirements (A$_{eff}$~=~0.6 m$^2$, 
HEW~=~10")}. 
\label{ss_table1}
\end{table}

\section{The young stellar population and the recent star formation history in the Galaxy}

The reason why X-ray observations are crucial for the study of the nearby, less than 10$^9$ yr old, stellar populations is the
well known fact that the X-ray luminosity decays by 1000-10000 times evolving from the PMS to the solar age; this decay occurs mainly during the main sequence phase. Such a behaviour is due to the decrease of angular momentum and rotational velocity with age. The younger, faster rotating, stars have a stronger magnetic field resulting in a higher X-ray coronal emission. For the sake of the reader let me remember that due to their fast rotation the older synchronized, so called active, binaries are also characterized by an intense coronal emission. While the age decay is the major observable effect we have also observed a softening of emitted X-ray spectrum with stellar age (cf. 
\citet{2003AN....324...77M} and reference therein cited).

\begin{figure*}[t]
 \centering
 \vspace*{0.3cm}
 \includegraphics[width=11.0cm]{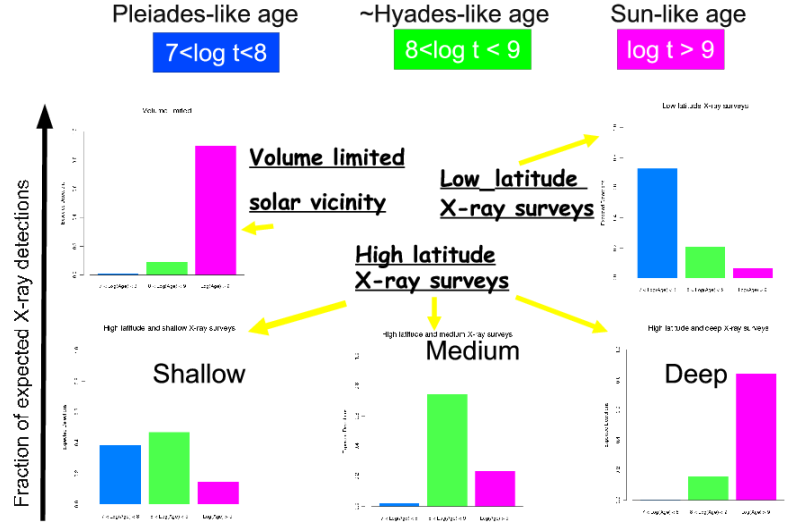}
 \caption{A schematic representation of the fraction of predicted X-ray normal star counts 
 for various directions and limiting flux for 3 distinct stellar age ranges (identified by distinct colours). 
In this illustrative calculation the X-ray luminosity of all stars for each given age range has been taken equal 
to the median L$_X$ of the corresponding sample. The absolute number of predicted stellar counts is different in each of the 5 panels.}
\label{fig0_ss}
\end{figure*}

Since the stellar X-ray (coronal) emission depends strongly on age, X-ray surveys play a key role in deriving i) the spatial distribution, in particular the densities and scale heights, of the young ($<$ 10$^9$ yr) stellar populations as well as ii) the star formation rate in the last billion year. Indeed in this age range optical photometric surveys are blind since there are no discernible color changes in stellar luminosity and one has to resort to spectroscopically discernible changes (i.e. the intensity of Lithium line) observed in dG-dK stars. Moreover, for dG stars this change mainly occurs after the first $\sim$ 5 x 10$^7$ yr.

\begin{figure*}[t]
 \centering
 \includegraphics[width=6.3cm,trim=0 30 0 30]{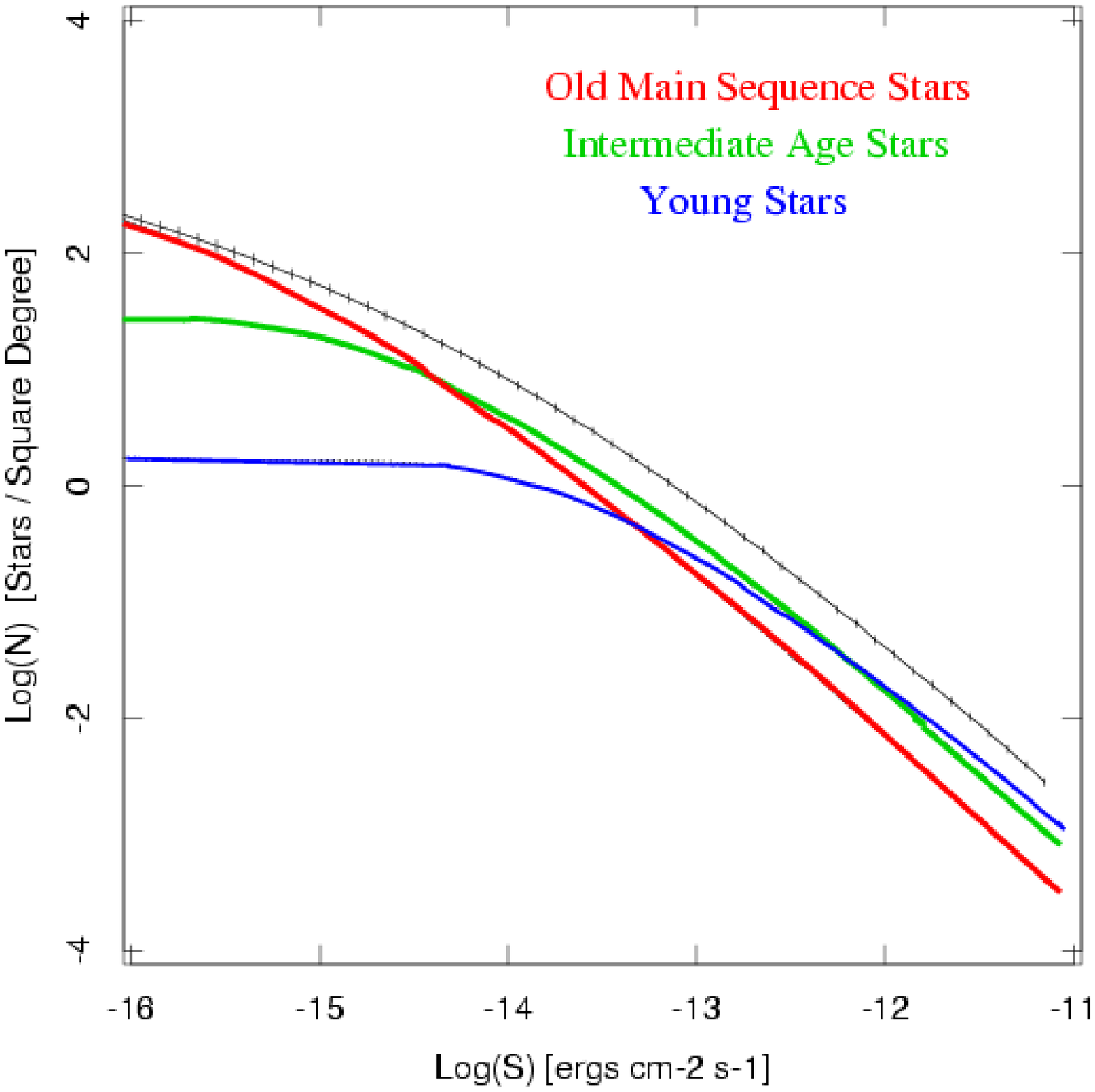}
 \includegraphics[width=6.3cm,trim=0 30 0 30]{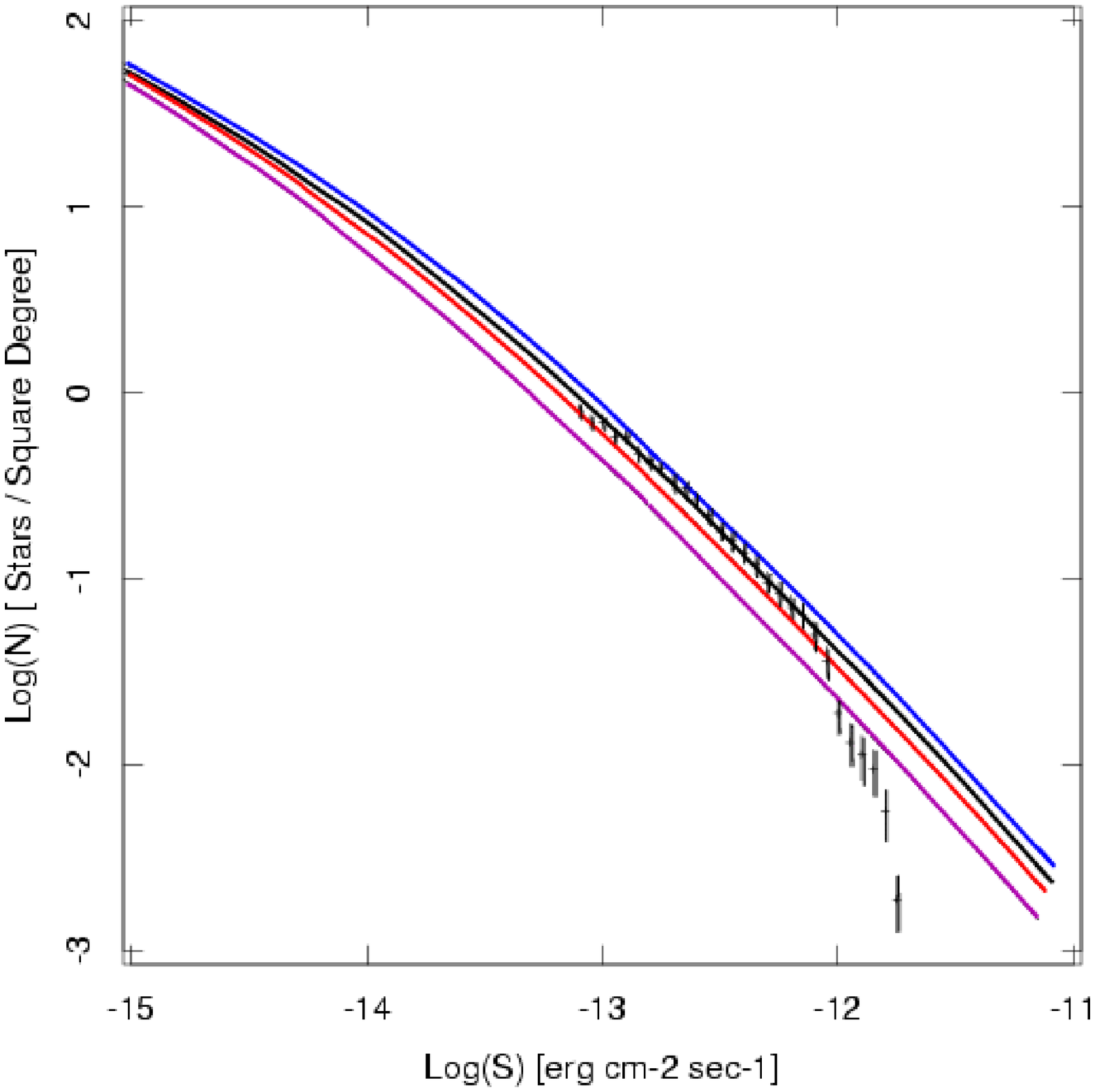}
 \caption{Left panel: X-COUNT predicted stellar number counts at high galactic latitude assuming a constant stellar birthrate for 3 distinct age ranges: young
stars (blue line), intermediate age stars (green line), and old
stars (red line), summed over ages (black line).
Right panel: observed (points \& error bar from EMSS data) and predicted 
Log(N)-Log($S_x$) toward the direction $l=90^\circ$ and $b=90^\circ$ 
assuming $\tau = \infty$ (black line), $\tau = 15~Gyr$ (red line), $\tau 
=-15~Gyr$ (blue line), and $\tau=5~Gyr$ (magenta line), [figures adapted from \citet{1993ApJ...412..618M}]. }
\label{fig1_ss}
\end{figure*}

In the "soft" ($\sim$ 0.2-10 keV range) X-rays we can observe young stars at much larger distances than old stars; hence young stars dominate shallow stellar X-ray selected sample while old stars dominate deep high latitude X-ray selected stellar samples (cf. Fig. \ref{fig0_ss}).
Active binaries, with their high X-ray luminosities, are selected as well, so companion optical data can be required to disentangle the case. Starting from a "classical" model \citep{ss_Bah80, ss_Bah86} for the stellar counts in the Galaxy  \citet{1992A&A...256...86F} have build a model, X-COUNT, that provides stellar counts in the X-ray band-pass. The model is based on i) average stellar spatial distributions for various age ranges, ii) an average spatial model of interstellar gas (and resulting extinction in X-rays), and Maximum Likelihood X-ray luminosity distribution functions from age-selected well studied open clusters and stellar samples. Using X-COUNT (or a similar model developed by \citealt{1996A&A...316...89G}) it is possible to compare observed and predicted X-ray stellar counts and, as a result, to derive the spatial distributions of stellar populations in the Galaxy. Following a seminal paper \citep{1988ApJ...324.1010F} on the stellar content of the {\it Einstein} Medium Sensitivity Survey \citep{ss_Gio90, ss_Sto91}, a detailed investigation of the stellar content of the {\it Einstein} EMSS (215 stars at a limiting f$_X \sim$ 2 x 10$^{-13}$ erg/sec/cm$^2$ over about $\sim$ 778 sq. deg.) has shown the presence of an excess of yellow (dG-dK) stars with respect to predictions \citep{1995A&A...296..370S}. 
This excess can be either young star or active binaries; an extensive optical campaign has shown that most of them are indeed young stars \citep{1993A&A...277..428F, 1995A&A...295..147F}. By comparing the observed and predicted stellar logN - LogS it is possible to infer the behaviour of the recent star formation rate in the Galaxy. This has been firstly demonstrated by \citet{1993ApJ...412..618M} who, using the EMSS data, have been able to show that some hypothetical temporal behaviours of the recent star formation rate are not consistent with available data (cf. Fig. \ref{fig1_ss}). 

The EMSS is by construction an high (b $>$ 20 deg) latitude survey, while the stellar density in the Galaxy is higher near the plane. In order to study the Galactic Plane content at a limiting flux, f$_X \sim 10^{-14}$ erg/sec/cm$^2$, 10 times deeper than the RASS, a ROSAT pointed Galactic Plane survey covering $\sim$ 2.5 sq. deg. was performed. The analysis of the 93 stars found in this survey \citep{2001MNRAS.326.1161M} has allowed us to conclude that the density of active stars in the Galactic Plane is larger than assumed in X-COUNT; henceforth either the star formation rate is increased in the last billion year or the young population scale heights are smaller than assumed in X-COUNT.

Coming back to high latitude surveys, the analysis of the stellar content (152 stars over 9 sq. deg.) of the ROSAT NEP survey (f$_X \sim 10^{-14}$ erg/sec/cm$^2$) has confirmed, but at one dex deeper limiting flux, the EMMS results, namely: for dA-dF and for dM X-COUNT predictions agree well with the observations, while a significant excess of yellow (dG-dK) stars is present 
\citep{2007A&A...461..977M}. \citet{2007A&A...463..165L} have analyzed the stellar content of the XBSS (XMM-Newton Bright Serendipitous Survey) confirming an excess of yellow (G+K) stars. A model calculation with a decreasing stellar birthrate is ruled out by observations. A constant SFR can reproduce the number of A and M stars, but underestimates the total number of observed stars. An increasing birthrate seems to work better: the total number of predicted sources agrees with the observations although it overestimates the total number of M stars. Those however can be hidden in some "yellow" binary.
The FGK star excess cannot be reproduced by using only a smooth stellar birthrate, unless the discrepancy between observations and predictions is due to a stellar population not (yet) included in the X-COUNT model. In summary the discrepancy between predicted and observed spectral type distributions could either be due to an "additional" population of young stars with small scale height or to a number of M dwarfs in binary systems with a yellow primary.

\begin{figure}[t]
 \centering
 \vspace*{0.3cm}
 \resizebox{\hsize}{!}{
 \includegraphics[width=11.2cm]{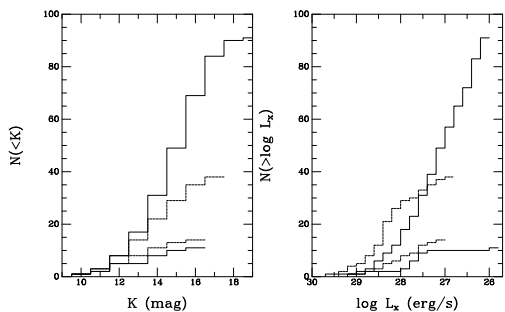}}
 \caption{Left panel: 
Comparison of cumulative distributions of stellar parameters for XCOUNT models of the CDF-N stellar population. Left: K-band magnitude. Right: X-ray luminosity. In each panel, histograms from top to bottom are as follows: total stellar population in the CDF-N field with V $<$ 22.5 and without X-ray selection (thin solid line), XCOUNT model prediction with standard settings including X-ray selection and no age $>$ 1 Gyr magnetic activity evolution (thin dashed line), XCOUNT model with rapid $age^{-2}$  X-ray decay (thick dashed line), and the observed distributions (thick solid line (figures from \citealt{2004ApJ...611.1107F}).
}
\label{fig2_ss}
\end{figure}

A contribution to this studies has been provided by the analysis of the stellar content of the Chandra Deep Field North \citep{2004ApJ...611.1107F}. With a quite small sample consisting of 11 stars with V $<$ 22.5, it has been possible to derive strong evidence of  
a decrease of X-ray luminosities over the 1Gyr $<$t$<$11Gyr age interval. With no decrease 39 rather than 11 stars should have been detected. The deduced "best fit" model has L$_X \sim age^{-2}$ ergs s$^{-1}$, which is faster than the $age^{-1}$ behaviour expected on the basis of known rotational spin-down rates and X-ray-activity relations (cf. Fig. \ref{fig2_ss}). 

However larger stellar samples (and deeper surveys) are required to firmly derive the recent star formation history, and specifically, the alternation of peaks and lulls of the star formation rate. This is an area in which WFXT can provide the still lacking data.

\section{The Gould Belt (or Disk?) nature}

\begin{figure}[t]
\centering
\resizebox{\hsize}{!} {
\includegraphics[width=7cm,angle=0]{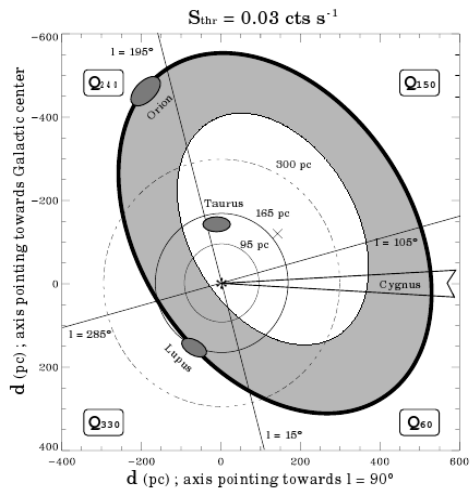}}
\caption{A sketch of the Gould Belt geometry projected on the Galactic plane.
The Gould Belt is assumed to be an ellipsoidal shaped ring with
semi-major and minor axes equal to 500 and 340 pc, respectively. The members 
are assumed to be located near the outer edge of belt (solid thick curve). The asterisk marks the
Sun position. The
depicted circles of radius 95, 165 and 300 pc centered on the Sun show
the X-ray horizon, at a PSPC limiting count-rate of 0.03 cts s$^{-1}$ ($\sim$ 3 x 10$^{-13}$ erg/s/cm$^2$),
for stars with log(L$_X$ [erg/s]) = 29.5, 30.0 and 30.5, respectively. The horizon for a G5 ZAMS star at
the Tycho completeness threshold (10.5 mag) is 160 pc. The grey shaded area
illustrates the alternative picture of the Gould Disk whose
members are spatially distributed between the inner and outer rings (figure from \citealt{1998A&A...337..113G}).}
\label{fig4_ss}
\end{figure}

\begin{figure}[h]
 \centering 
 \vspace*{0.3cm}
 {\hspace{-0.5cm}\includegraphics[width=7.0cm,angle=0]{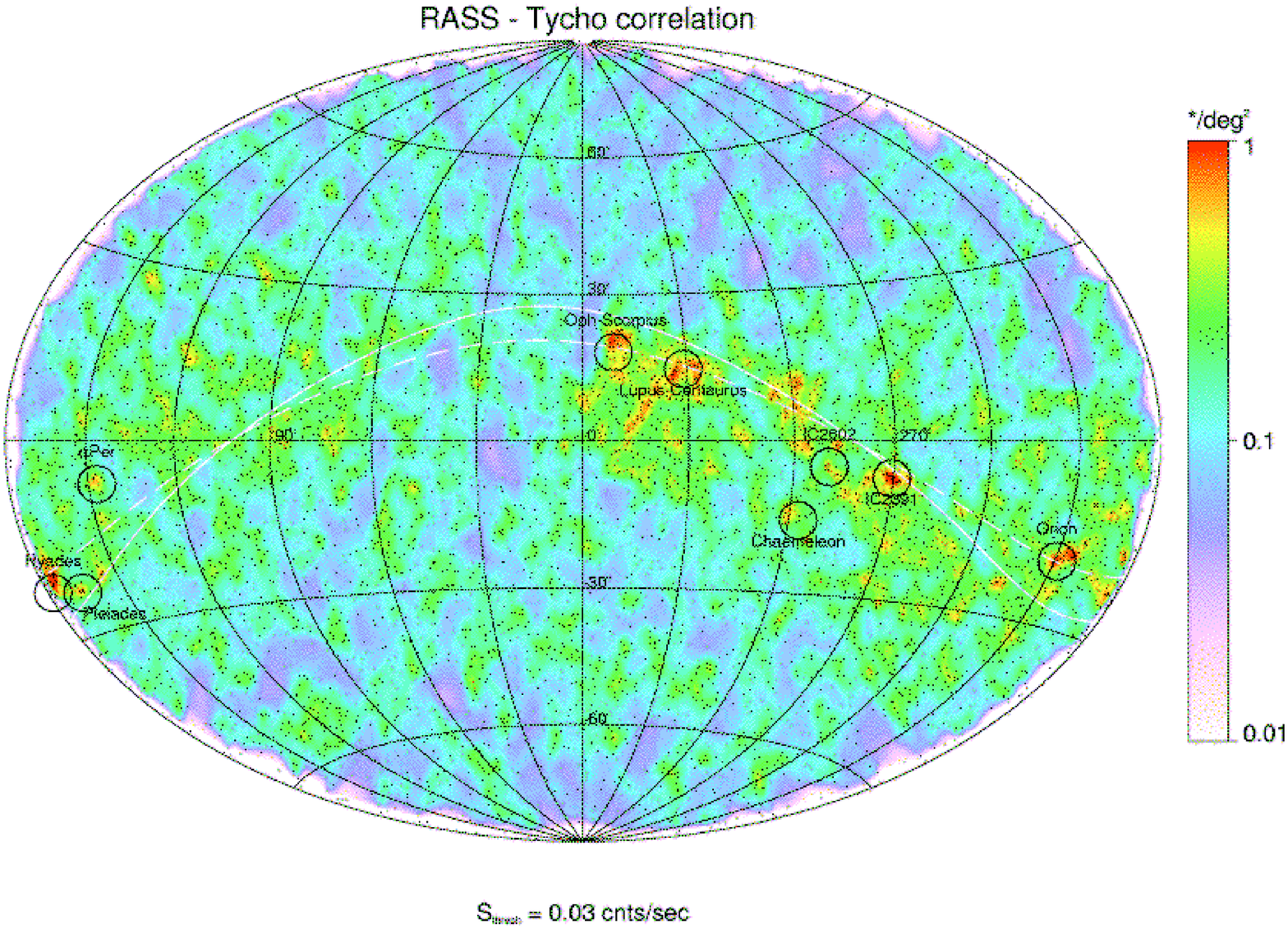}
 \includegraphics[width=3.6cm,height=5.5cm,angle=90]{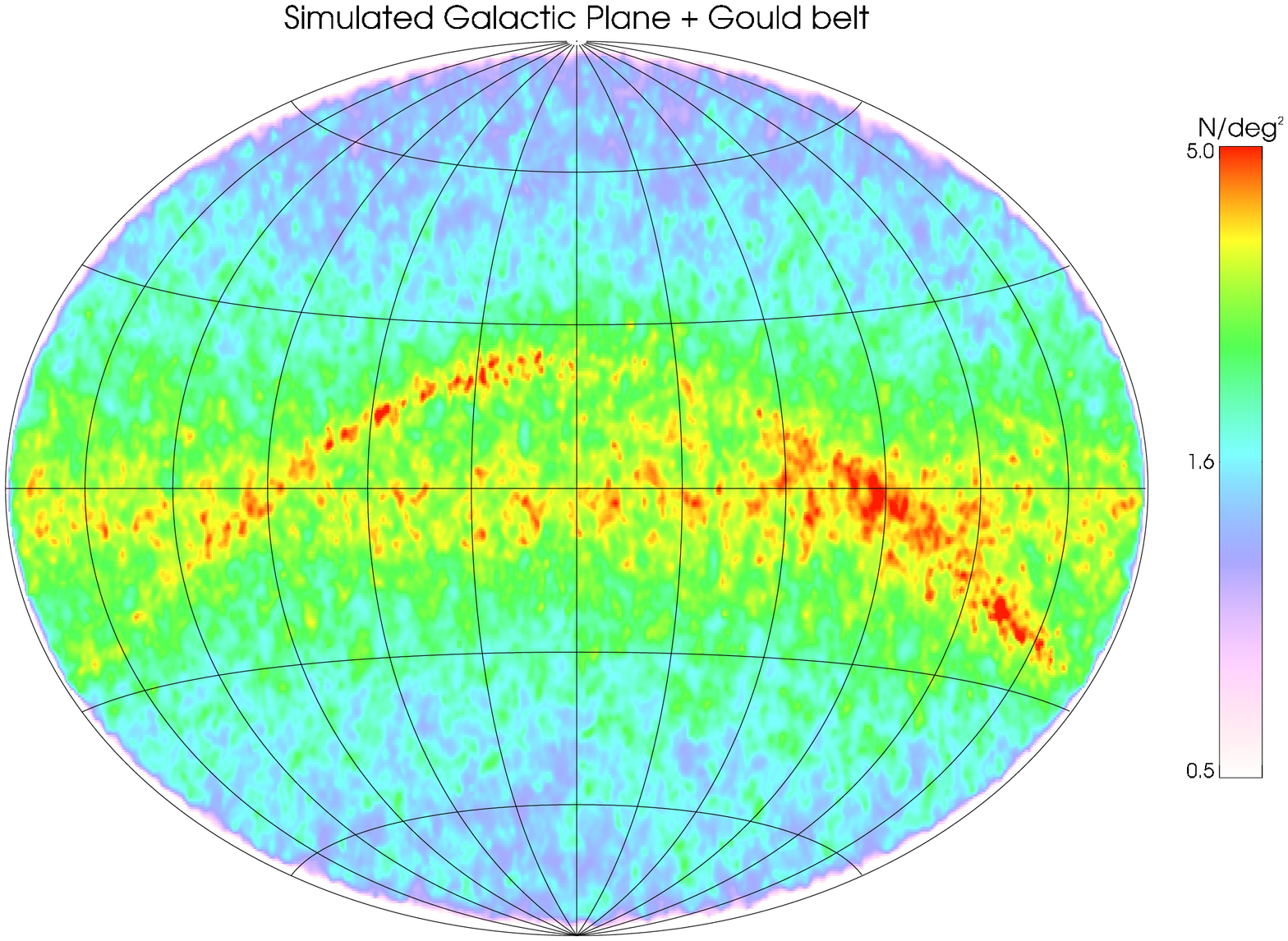}}
 \caption{Top panel: All sky distribution in galactic coordinates of RASS-Tycho stars (black dot) at a PSPC count-rate threshold S = 0.03 cts s$^{-1}$ ($\sim$ 3 x 10$^{-13}$ erg/s/cm$^2$). Color codes the surface density, the enhancement at low
galactic latitude is clearly visible. The dashed line indicates Gould belt. Black circles indicate young
clusters and star forming regions that show up in the RASS-Tycho sample. Bottom Panel:
A simulated all sky distribution in galactic coordinates of X-ray emitting stars at a limiting flux of ~ 5 10$^{-14}$ erg s$^{-1}$ cm$^{-2}$, 30-50 times deeper than the RASS. Color codes the surface density. At this deeper flux the Gould Belt, if a physical structure, will clearly be discernible (Simulation courtesy of P. Guillout). 
}
\label{fig3_ss}
\end{figure}

\citet{1998A&A...334..540G} have performed the positional cross-identification between the RASS source list (at the limiting count rate of S~=~0.03 PSPC c/s) and the stars listed in the Tycho catalog. The large scale sky distribution of the X-ray emitting Tycho stars (cf. Fig. \ref{fig3_ss}) shows an enhancement (apart around the Galactic Plane) that has been interpreted as a structure of young nearby stars, likely, coincident with the so called Gould Belt. The available data have still left open various questions: i) Is this enhancement due to a real, recently formed, physical structure ?, ii) Is this structure a Belt or instead is more similar to a Disk ?; iii) Is this structure related to nearby star formation processes and some "local" triggering mechanism(s) ? 
With the RASS we have been able just to see the side of a hypothetical belt/disk structure nearest to the Sun, while the farthest side is beyond the RASS horizon (cf. Fig. \ref{fig4_ss}). To answer those and other connected questions we definitively need a much deeper large area X-ray survey with a spatial resolution better than XMM-Newton (to select the possible members of such a structure) and we need the GAIA data to derive distances and 3-d space velocities to discriminate the very likely members. A joint WFXT/GAIA investigation will provide an invaluable contributions to assess the nature of the Gould Belt/Disk and more generally of the star formation process in the solar neighborhood. This is clearly shown in pictorial form by the simulation (cf. Fig. \ref{fig3_ss}) of the outcome of a 30-50 times deeper than the RASS X-ray survey with an angular resolution of 10". The nature of the Gould Belt (or Disk) will clearly be derived from a properly selected survey of a (large) sky area, without the need to an all sky coverage. With an efficency of 80\% a dedicated medium depth survey (cf. \ref{ss_table1}) of about 200 sq. deg. will require about 1-2 months.  

\section{Formation and evolution of Young Stellar Clusters (YSC) and their IMFs}

The study of the formation and evolution of the YSC (Young Stellar Clusters) and of their IMF's (Initial Mass Function) is another wide research theme on which a mission like WFXT will provide otherwise impossible to obtain data. 
For a thorough discussion of this theme I point the reader to a recent white paper by \citet{2009astro2010S..77F} submitted to the US Decadal Committee lead by R. Blanford.

Let me just recall the major open issues and related questions: th

Do clusters form rapidly during a single collapse event or slowly over many crossing times ?
Why are massive stars rare ? Do they form via accretion disks or stellar mergers ?
How does the feedback from OB stars both halt and promote further star formation?
Is the stellar IMF truly universal over a wide range of cloud conditions, and what produces its distinctive shape ? 
When and why does primordial mass segregation (if any) occur ?
What effect do shocked OB winds have on the physics of the HII region and the confining GMC ? 
What fraction of stars in the Galaxy form from triggered processes?
What determines whether a YSC survives the dispersal of its parental molecular gas and becomes a bound open or globular cluster?
How does the cluster environment influence the evolution of protoplanetary disks and subsequent formation of planetary systems?

Such a research theme definitively requires panchromatic studies: optical, infrared (from ground and space), millimeter (ALMA), as well as X-rays observations are needed. Given the typical extent of relevant targets (that often cover up to tens of square degrees) a WFXT is a key element of the needed instrumental suite. While there are many reasons for the key role played by
X-ray observations, let me remember a very basic one: IRAC Spitzer data as well as the new data from Herschel are having an impressive impact in the field, but a reliable IMF can hardly be derived from Spitzer data alone since one will very likely miss a large fraction of WTTs. For example, the identification of a complete sample of members (both Classical T Tauri and Weak Line T Tauri) has proven to be a very crucial point in the investigations of the "environmental" effects on disk evolution (cf. \citealt{2007A&A...462..245G, 2009A&A...496..453G}).

\section{Continuous monitoring: Variability of Class I e Class II emission} 

Recently Favata and collaborators have obtained a 23-days long uninterrupted COROT observation of the Young Cluster/SFR NGC 2264 (with an age of $\sim$ 3 Myr). This was a specific additional program aimed to study the variability phenomena of a large sample of NGC~2264 members. The long photometric series are allowing many investigations, in particular to look for light curve variability of the same nature as that observed in the classical T Tauri star AA Tau \citep{ss_Alencar10}. This variability was interpreted as due to inner warp dynamic associated to and controlled
by the interaction between stellar magnetic field and the inner disk region . The classification purely based on the analysis of light curves is in good agreement with the Spitzer IRAC classification systems, about 40\% of of NGC~2264 members show warped disks, i.e. evidence of an intense accretion process mediated by a highly dynamical star-disk magnetospheric interaction. If this picture is correct then we should expect evidence of time correlation between optical and X-ray variability among the sample of AA Tau-like PMS stars; in particular, we should see variability in the emitted intensity as well as in the emitted spectrum (mostly related to a time variable absorption).

Dedicated long observations, such as those of the Chameleon SFR that will be possible with the ESA CV proposed PLATO mission, together with simultaneous (week) long continuous observations with WFXT will definitively provide a big step forward in our understanding of the disk-star interaction and on the role of magnetic field. This would require a couple months of dedicated WFXT observations.   
 
\section{Long Term Programs: X-ray Cycles of late-type stars}

\begin{figure}
 \centering
 \resizebox{\hsize}{!}{
  \includegraphics[width=6cm]{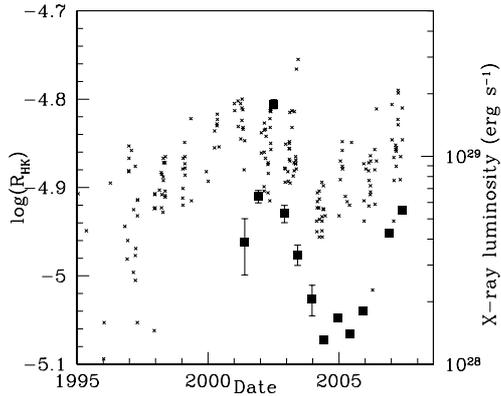}}
 \caption{The variation of the X-ray luminosity for HD 81809
as observed by XMM-Newton since 2001 plotted together with the CaII
data available until 2007.}
\label{fig5_ss}
\end{figure}

Currently the WFXT mission concept that has been proposed to the US Decadal Survey considers a mission lifetime of 5 years. However based on the experience of other highly successful X-ray observatories (like Chandra and XMM-Newton), whose lifetime has been extended to more than a decade, I think that we have to consider from the beginning the possibility to include few selected long-term plan investigations that ask for repeated observations over a decade. As an example, let me focus on the case of X-ray cycles of quiet late-type stars, whose X-ray luminosity is similar to that of our Sun, L$_X$ $\sim$ 10$^{27}$ erg/sec. While for our Sun the evidence of a long term (11 years) cycles of the emission in many bands including the X-ray one is based on firm observational evidence, a similar evidence was completely lacking in the case of other X-ray quiet stars; indeed the spotted nature of the available observations had been a serious observational limitation. It has been only for the perseverance of some of us (and good will of many TAC members) that, thanks to a decennial campaign of XMM-Newton repeated observations, we have been able to provide in a couple of cases clear observational evidence of X-ray cycles: probably the best case, HD 81809 \citep{2008A&A...490.1121F} is shown in Fig. \ref{fig5_ss}. I guess that a proper planned campaign of WFXT observations can enlarge this very small sample and allow us to verify if X-ray cycle and CaII periods do, in general, agree or not and, more generally, what are the conditions under which X-ray cycles may occur. While this is just an example, I feel it contains a lesson: even if formally a mission, WFXT in such a case, has a nominal lifetime too short for a decade-long investigation (this was indeed the case of XMM-Newton), it is definitively worth to start few selected of such long-term investigations, since they can provide scientific returns otherwise impossible. 

\section{Concluding Remarks}

WFXT surveys will allow deriving the properties of $< 10^9$ yr old population in the Galaxy. Properly planned follow-up optical observations are needed.
The shallow, medium and deep planned high latitude WFXT surveys will allow determining densities and scale heights of young, intermediate and old stellar populations and the history of the star formation rate in the last billion year,
a period of time that optical surveys alone cannot explore. 
A properly planned, dedicated survey at low-galactic latitude will allow investigating the nature and origin of the Gould Belt/Disk and of its low-mass stellar population, i.e. the nature of one of the recent episodes of star formation occurred in the solar neighborhood. 
A WFXT mission can step forward our knowledge on many open issues on the physics and process at work in Young Stellar Clusters (and associated proto-planetary systems) formation and early evolution. 
Key laboratories are the nearby ($<$ 1 kpc) SFRs dispersed on a large area
of sky. Such investigation will greatly take advantage of the $\sim$ 5" angular resolution, the current WFXT goal. More in general stellar studies will require that some time will be devoted to surveying properly selected regions: medium and medium/deep observations are required in order not only to find the sources, but to ease the identification process by using the collected X-ray spectra. 

\begin{acknowledgements}
I acknowledge many enlightening discussions with G. Micela, E. Feigelson, 
E. Flaccomio and F. Damiani and the material they have kindly provided.
This work was partially supported from ASI/INAF Contracts I/088/06/00 and 
I/023/05/0.
\end{acknowledgements}

\bibliographystyle{aa}

\end{document}